\documentclass[conference]{IEEEtran}
\usepackage{cite}
\usepackage[]{graphicx}
\ifCLASSINFOpdf
\else
\fi
\usepackage{amsmath}
\usepackage{array}
\begin{document}

\title{Critical Current Oscillations of Josephson Junctions Containing PdFe Nanomagnets}%

\author{Joseph A. Glick, Reza Loloee, W. P. Pratt, Jr. and Norman O. Birge~\IEEEmembership{, Member,~IEEE}\\% <-this % stops a space
\IEEEauthorblockA{Dept. of Physics and Astronomy, Michigan State University, East Lansing, MI 48824, USA\\
Email: birge@pa.msu.edu\\
Manuscript received September 5, 2016}}

\maketitle

\begin{abstract}
Josephson junctions with ferromagnetic layers are vital elements in a new class of cryogenic memory devices. One style of memory device contains a spin valve with one ``hard'' magnetic layer and one ``soft'' layer.  To achieve low switching fields, it is advantageous for the soft layer to have low magnetization and low magnetocrystalline anisotropy.  A candidate class of materials that fulfills these criteria is the Pd$_{1-x}$Fe$_{x}$ alloy system with low Fe concentrations.  We present studies of micron-scale elliptically-shaped Josephson junctions containing Pd$_{97}$Fe$_{3}$ layers of varying thickness. By applying an external magnetic field, the critical current of the junctions are found to follow characteristic Fraunhofer patterns. The maximum value of the critical current, extracted from the Fraunhofer patterns, oscillates as a function of the ferromagnetic barrier thickness, indicating transitions in the phase difference across the junction between values of zero and $\pi$.
\end{abstract}


% \fi
%\IEEEpeerreviewmaketitle

\section{Introduction}
Josephson junctions containing ferromagnetic (F) layers are being studied by many researchers to create an energy-efficient, fast, non-volatile memory for superconducting computing~\cite{Bell2004, Ryazanov2012, Larkin2012, Vernik2013, Qader2014, Baek2014, Gingrich2016}. Recently our group demonstrated that a phase-controllable memory element can be made from a Superconducting QUantum Interferance Device (SQUID) containing two Josephson junctions with the structure S/F$^{\prime}$/N/F$^{\prime \prime}$/S, where S is a superconductor, F and F$^{\prime}$ are ferromagnetic materials, and N is a normal metal~\cite{Gingrich2016}. In this and other similar proposals~\cite{Bell2004, Qader2014, Baek2014}, one of the ferromagnetic layers (F$^{\prime \prime}$, the ``free layer'') can be made to switch it's magnetization direction to be parallel or anti-parallel to the other layer (F$^{\prime}$, ``hard layer'') by application of a small magnetic field. The thicknesses of the F$^{\prime}$ and F$^{\prime \prime}$ layers are set so that when their magnetization vectors are parallel the junction is in the $\pi$-phase state, and when the two layers are anti-parallel the junction is in the 0-phase state, as dictated by the superconducting proximity effect.

To maximize energy-efficiency in a memory application, it is desirable to use a free layer whose magnetization direction can be controllably switched by a very low applied field. The magnetic material used for the free layer should thus have low magnetization and low magnetocrystalline anisotropy.  Here we study the properties of the soft magnetic alloy Pd$_{97}$Fe$_3$, which is under consideration for the free layer material. Dilute PdFe alloys have been known for several decades to have very low magnetocrystalline anisotropy \cite{Senoussi1977}, and our own previous work on the alloy with 1.3\% Fe concentration found it to have a spin diffusion length of 9.6 $\pm$ 2 nm ~\cite{Arham2009}.  Josephson junctions containing PdFe with a lower Fe concentration of $\approx$1\% have already been studied by other groups~\cite{Ryazanov2012, Larkin2012, Vernik2013} with an eye toward applications in cryogenic memory.  We have tried using Pd$_{98.7}$Fe$_{1.3}$ as the free layer in controllable spin-triplet Josephson junctions~\cite{GingrichThesis2014}, but the results were not satisfactory.  That work provided the main motivation for studying PdFe alloys with somewhat higher Fe concentrations.

\section{Sample Fabrication and Characterization}
We first characterized the magnetic properties of unpatterned continuous Pd$_{97}$Fe$_3$ films via SQUID magnetometry. Thin films of Nb(5)/Cu(5)/PdFe(d$_F$)/Cu(5)/Nb(5), with thicknesses in nanometers, were deposited via dc sputtering in an Argon plasma with pressure 1.3 $\times$ 10$^{-3}$ Torr. Prior to sputtering the base pressure of the chamber was 2$\times$10$^{-8}$ Torr. During the deposition the sample temperature was held between $-30\,^{\circ}\mathrm{C}$ and $-20\,^{\circ}\mathrm{C}$. The thicknesses of the various deposited materials were controlled by measuring the deposition rates (accurate to $\pm 0.1 \mathrm{\AA}$/s) using a crystal film thickness monitor.

\begin{figure}
	\begin{center}
		\includegraphics[width=2.8 in]{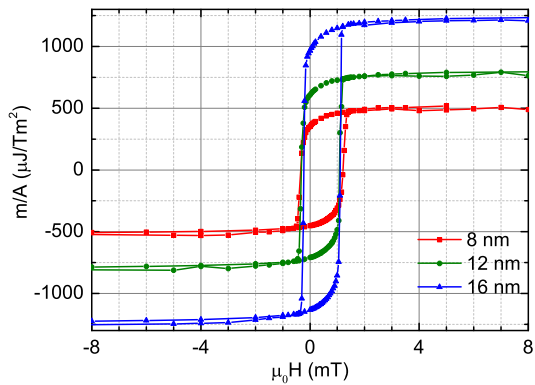}	
	\end{center}
	\caption{\label{fig:PdFeSQUIDMagnetomemer} Hysteresis loops of unpatterned films containing PdFe with thickness $d_{\mathrm{PdFe}}$ spanning 8-16 nm. Plotted is the magnetic moment divided by the sample area versus the applied field, measured using SQUID magnetometry. For the three samples the magnetization is approximately constant, $M=$ 90 $\pm$ 9 kA/m. The data are slightly shifted along the field axis due to a small amount of trapped flux in the solenoid of the SQUID magnetometer. From the data we extract a magnetic dead layer thickness of $d_{\mathrm{dead}}$=2.8 $\pm 0.9$ nm, discussed in the text.}
\end{figure}

The samples were measured using a Quantum Design SQUID magnetometer at 5 K, with the applied magnetic field parallel to the film plane. The hysteresis loops of films with $d_\mathrm{PdFe}$ = 8-16 nm are shown in Fig.~\ref{fig:PdFeSQUIDMagnetomemer}. The saturation magnetization per unit volume is nearly constant for the three samples. Plotting the saturation magnetic moment divided by the sample area versus $d_\mathrm{PdFe}$ and fitting to a straight line gives a slope which corresponds to a magnetization of $M=$ 90 $\pm$ 9 kA/m. Meanwhile, the x-intercept shows a magnetic dead layer thickness of $d_{\mathrm{dead}}$=2.8 $\pm 0.9$ nm. Note that these unpatterned films contain many magnetic domains so that the switching mechanism is governed by domain-wall motion; hence the film results should not be directly compared to the switching behavior of the nanomagnets in our SFS junctions, discussed later.

\begin{figure}
	\begin{center}
		\includegraphics[width=2.2 in]{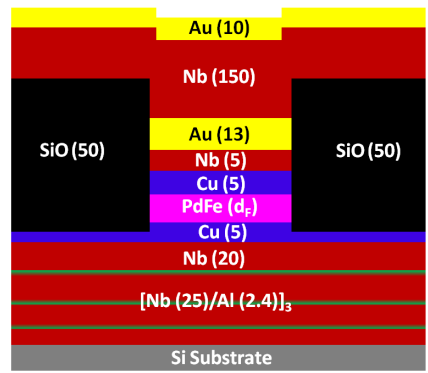}
	\end{center}
	\caption{\label{fig:SampleStructure} A schematic showing the vertical cross-sectional structure of our SFS Josephson junctions. The Pd$_{97}$Fe$_{3}$ thickness $d_F$ ranges from 9 to 36 nm. All thicknesses are given in nanometers.}
\end{figure}

In a separate sputtering run we fabricated SFS Josephson junctions containing PdFe using the same techniques described above. Prior to sputtering, photolithography was used to define the geometry of our bottom wiring layer, which consists of the sequence [Nb(25)/Al(2.4)]$_3$/Nb(20)/Cu(5)/PdFe(d$_F$)/Cu(5)/Nb(5)/Au(15), and was sputtered without breaking vacuum. A schematic of the full sample structure is shown in Fig.~\ref{fig:SampleStructure}.

To achieve sharp magnetic switching we grew the ferromagnets on a smooth [Nb/Al] multilayer used in previous works~\cite{Wang2012, Thomas1998, Kohlstedt1996}. Using atomic force microscopy (AFM), the roughness of the [Nb/Al] multilayer we independently measured was $\approx 2.3$ $\mathrm{\AA}$, which is smoother than our sputtered Nb(100) films with roughness $>5$ $\mathrm{\AA}$. A 5 nm Cu spacer layer was used on either side of the ferromagnet to improve it's magnetic properties and the samples were capped with a thin layer of Nb and Au to prevent oxidation.

The elliptically-shaped junctions were patterned via electron-beam lithography followed by ion milling in Argon, with the same process used our previous work~\cite{Niedzielski2015, Glick2016}.  The junctions have an aspect ratio of 2.5 and area of 0.5 $\mu$m$^2$, which is small enough to make some magnetic materials, such as NiFe and NiFeCo, single domain~\cite{Glick2016}.

Outside the mask region, ion milling was used to etch through the capping layer, the F layer, and half-way into the underlying Cu spacer layer. After ion milling, we thermally evaporated a 50 nm thick SiO layer to electrically isolate the junction and the bottom and top wiring layers. During the ion milling and SiO deposition, to prevent the e-beam resist from over-heating, the back of the substrate was pressed against a Cu heatsink coated with thin layer of silver paste to improve thermal contact.

Finally, the top Nb wiring layer was patterned using similar photolithography and lift-off processes as the bottom leads. Residual photoresist was cleaned from the surface of the samples with oxygen plasma etching followed by \textit{in-situ} ion milling in which 2 nm of the top Au surface was etched away prior to sputtering. The sputtered top electrode consists of Nb(150 nm)/Au(10 nm), ending with Au to prevent oxidation.

\section{Measurement and Analysis}
The samples were wired to the leads of a dip-stick probe using pressed indium solder and inserted into a liquid-He dewar outfitted with a Cryoperm magnetic shield. A superconducting solenoid on the dipping probe is used to apply uniform magnetic fields along the long-axis of the elliptical junctions over a range of \mbox{-60} to 60 mT. The current-voltage characteristics of the junctions were measured in a standard four-terminal configuration at 4.2 K. The I-V curves were found to have the expected behavior of overdamped Josephson junctions~\cite{Barone1982},
\begin{equation}
\label{eqn:SquareRootFunction}
V= R_N \sqrt{I^2 - I_c^2}, \hspace*{0.1in} I \ge I_c,
\end{equation}
where $I_{c}$ is the critical current and $R_N$ is the sample resistance in the normal state. $R_N$ is the slope of the linear region of the I-V curve when $|I| \gg I_c$, and was independently confirmed using a lock-in amplifier. Measurements of the area-resistance product in the normal state yielded consistent values of $AR_N$ = 11 $\pm$ 1 f$\Omega$-$m^2$, an indicator of reproducible high quality interfaces. 

\begin{figure}
	\begin{center}
		\includegraphics[width=3.2 in]{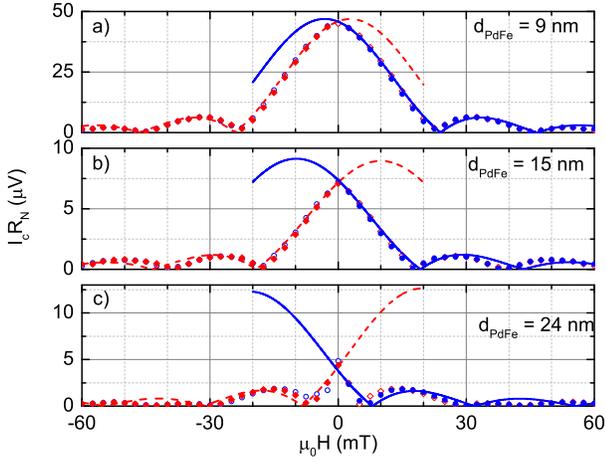}			
	\end{center}
	\caption{\label{fig:PdFe_Fraunhofers} PdFe Fraunhofer patterns:  Critical current times the normal state resistance, $I_c R_N$, is plotted versus the applied field $H$, for three samples with $d_{\mathrm{PdFe}}$ equal to (a) 9 nm, (b) 15 nm, and (c) 24 nm. The data before $H_{\mathrm{switch}}$, the field at which the PdFe magnetization vector reverses direction (solid markers), and the corresponding fits (lines) to Eqn.~\ref{eqn:FraunhoferAiryFit} show good agreement for both the positive (red, dashed) and negative (blue) field sweep directions. The hollow circles are the corresponding data points after $H_{\mathrm{switch}}$. The Fraunhofer patterns display magnetic hysteresis and are increasingly shifted with larger $d_{\mathrm{F}}$.}
\end{figure}

``Fraunhofer'' diffraction patterns, shown in Fig.~ \ref{fig:PdFe_Fraunhofers}, were obtained by plotting $I_c R_N$ as a function of the applied magnetic field. The expected functional form of the Fraunhofer pattern for elliptical junctions is an Airy function~\cite{Barone1982},
\begin{equation}
\label{eqn:FraunhoferAiryFit}
I_{c}=I_{c0} \left| 2 J_{1} \left( \pi \Phi / \Phi_{0} \right) / \left( \pi \Phi / \Phi_0 \right) \right|,
\end{equation}
where $J_1$ is a Bessel function of the first kind, $I_{c0}$ is the maximum critical current, and $\Phi_0 = h/2e$ is the flux quantum. The magnetic flux through the junction is given by \cite{Khaire2009}\footnote{We correct a missing factor of $\mu_0$ in the corresponding equation in Ref.~\cite{Khaire2009}},
\begin{equation}
\label{eqn:magneticflux}
\Phi=\mu_0 H w (2 \lambda_L+2d_N+d_F) + \mu_0 M w d_F,
\end{equation}
where $H$, $w$, $d_N$ and $d_F$ are the applied field, the junction width, and the thicknesses of the normal metal and F layer, respectively. $\lambda_L$ is the London penetration depth of the Nb electrodes, which we keep fixed at 85 nm, as determined by data obtained in our group over many years \cite{Khaire2009}. The last term in Eqn.~\ref{eqn:magneticflux} describes the flux due to the magnetization $M$ of the nanomagnet, which is valid if $M$ is uniform and is oriented in the same direction as the applied field $H$. In Eqn.~\ref{eqn:magneticflux} we have omitted the much smaller flux terms from the uniform demagnetizing field and any magnetic field from the nanomagnet that returns between the top and bottom Nb electrodes. The Fraunhofer pattern will be shifted by an amount $H_{\mathrm{shift}}= -M d_F/(2 \lambda_L+ d_F+ 2 d_{\mathrm{Cu}})$ along the field axis due to Eqn.~\ref{eqn:magneticflux}.

The Fraunhofer patterns in Fig.~\ref{fig:PdFe_Fraunhofers} were collected by the following process: First we fully magnetized the nanomagnet with an applied a field of -60 mT, then ramped the field to +60 mT in steps of 2.5 mT, measuring $I_c$ at each step. 

The data follow the expected Airy function from the initialization field up to the beginning of a small field range, $H_{\mathrm{switch}} >$ 0, during which the ferromagnet switches the direction of it's magnetization vector. Beyond $H_{\mathrm{switch}}$ the data jump to another Fraunhofer pattern that is shifted in the opposite direction. To measure the magnetic hysteresis, as done in previous works~\cite{Baek2014,Niedzielski2015,Glick2016}, we then swept the applied field in the opposite orientation.  

The data prior to the magnetic switching event were fit to Eqn.~\ref{eqn:FraunhoferAiryFit} with $I_{c0}$, $w$, and $H_{\mathrm{shift}}$ as fitting parameters. In Fig.~\ref{fig:PdFe_Fraunhofers}, for both the positive (red) and negative (blue) sweep directions, the corresponding fits (lines) show excellent agreement with the data (solid markers). The hollow markers denote the data after $H_{\mathrm{switch}}$, and closely correspond to the Fraunhofer pattern in which the field is swept in the opposite orientation. The excellent nature of the Fraunhofer patterns allow us to extrapolate the maximum value of I$_c$, albeit with a larger uncertainty, even when the value of $H_{\mathrm{shift}}$ approaches the first minimum in the Airy function. The nodes in the Fraunhofer pattern nearly approach $I_c=0$, indicating a robust SiO barrier around the junction. The data typically follow the Airy function through zero field before the relatively sharp magnetic switching event, but for a few samples did not. Therefore, it is difficult to determine if the nanomagnets contain a single magnetic domain near zero field.  

\begin{figure}
	\begin{center}
		\includegraphics[width=3.0 in]{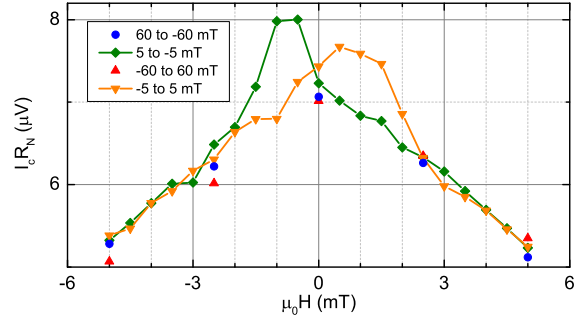}
	\end{center}
	\caption{\label{fig:15nmPdFe_ZoomedInFraunhofer} $I_c R_N$ is plotted versus the applied field $H$ for the same Josephson junction shown in Fig.~\ref{fig:PdFe_Fraunhofers}(b), zoomed-in on the central peak. Separate measurements using small initialization fields of $\pm$5 mT and finer step size (green and orange data points) show the behavior of the magnetic switching. The reversal of the PdFe magnetization direction for the two sweep directions begins at $H_{\mathrm{switch,1}}=$ 1.0 mT (orange) and -0.5 mT (green) and ends at $H_{\mathrm{switch,2}}$ = 2.5 mT (orange) and -2.0 mT (green). During the switching event the data deviate from the expected Fraunhofer pattern fit. As the field approaches $H_{\mathrm{switch,2}}$ the data converge with the corresponding measurements from Fig.~\ref{fig:PdFe_Fraunhofers}(b) where much larger $\pm$60 mT initialization fields were used (blue and red points). Lines connect the adjacent finer spaced data for clarity.}
\end{figure}

The switching characteristics of the PdFe layer were maintained even when smaller initialization fields were used. After returning the field to zero, we measured the Fraunhofer pattern again, sweeping the field from only $\pm$ 5 mT in both directions at finer field steps of 0.5 mT, as shown in Fig.~\ref{fig:15nmPdFe_ZoomedInFraunhofer} (green and orange points), where we have zoomed-in on the central peak. It is clear that the junctions switch the direction of their magnetization over a range of field values. To characterize the magnetic switching we use two parameters: $H_{\mathrm{switch, 1}}$, denoting the beginning of the switching event, is the field at which $I_c$ begins to deviate from the initial Airy function, and $H_{\mathrm{switch, 2}}$, denoting the end of the switching event, is the field at which $I_c$ joins the corresponding shifted Airy function. Across the range of thicknesses studied, on average the junctions began to switch at a very low field $|H_{\mathrm{switch,1}}|$ =  0.4 mT with standard deviation 0.6 mT, and completed the switching process at $|H_{\mathrm{switch,2}}|$ =  2.4 mT with standard deviation 0.9 mT. The value of $|H_{\mathrm{switch,1}}|$ for PdFe is smaller than found in Ni$_{81}$Fe$_{19}$-based junctions of similar construction measured by our group~\cite{Glick2016}, however $|H_{\mathrm{switch,2}}|$ is comparable. The low Fe concentration in the Pd$_{97}$Fe$_3$ alloy may give rise to this gradual switching behavior. Prior work on an alloy with lower Fe concentration, Pd$_{99}$Fe$_1$, showed that the ferromagnetic behavior of thin films are controlled by the presence of weakly coupled ferromagnetic clusters~\cite{Uspenskaya2013}.

Repeating the measurement at even lower initialization fields (3 mT) sometimes caused irregular and  irreproducible changes to $I_c$ and $H_{\mathrm{shift}}$. We surmise that too low of an initialization field allows domain walls to form within the junction, which disturb the magnetic switching. Hence, if Pd$_{97}$Fe$_3$ layers are used in cryogenic memory, an initialization field of at least 5 mT would be necessary to reproducibly magnetize the nanomagnet. 

\begin{figure}
	\begin{center}
		\includegraphics[width=3.0 in]{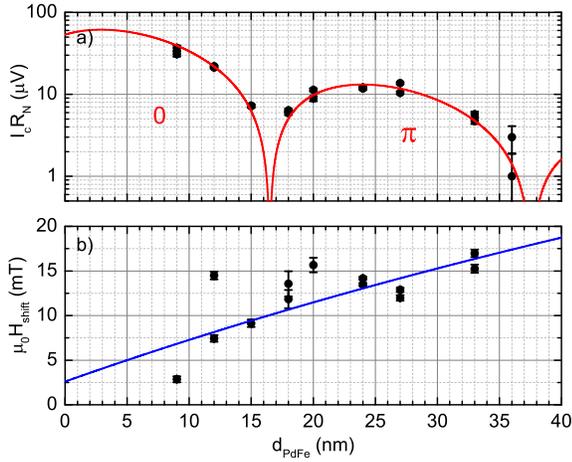}
	\end{center}
	\caption{\label{fig:IcRn_Hshift_PdFe} a) The maximal $I_c$ times R$_N$ is plotted versus $d_{\mathrm{PdFe}}$ for many samples, with the error bars determined by the goodness of fit parameters of the individual Fraunhofer patterns. The minima indicate the critical PdFe thicknesses at which the junctions transition between the 0 and $\pi$-phase states. The solid red line is a fit to the data using Eqn.~\ref{eq:IcRnZeroPiTransition}. b) The Fraunhofer pattern field shift $H_{\mathrm{shift}}$ increases with $d_{\mathrm{PdFe}}$. The blue line is the fit to Eqn.~\ref{eqn:Hshift}, which yields $M$ = 72 $\pm$ 16 kA/m and $d_\mathrm{dead}$ = -4 $\pm$ 5 nm.}
\end{figure}

In Fig.~\ref{fig:IcRn_Hshift_PdFe}(a) we plot $I_c R_N$ for many samples of varying ferromagnet thicknesses $d_F$. The junctions transition from a 0 to $\pi$-phase state at the value of $d_F$ where the first deep local minima occurs. In Fig.~\ref{fig:IcRn_Hshift_PdFe}(a) $I_c$ denotes the maximum critical current obtained from the Fraunhofer pattern fits. The 0 to $\pi$-phase state transition occurs at thicknesses of about $d_{F}$=16.5 nm, followed by a $\pi$-to-0 phase transition near $d_{F}$=38 nm. 

Theoretical predictions describe the behavior of $I_c R_N$ versus $d_F$ as an oscillating function with either an exponential decay for diffusive transport or an algebraic decay for ballistic transport~\cite{Buzdin_SFReview2005}. Robinson \textit{et al.} used the ballistic form to fit data from junctions containing very thin elemental ferromagnetic layers like, Ni, Co, and Fe~\cite{Robinson2006, Robinson2007}, but when grown thicker, the data were better modeled by the diffusive limit. In materials where the majority and minority spin bands have nearly identical properties, the diffusive limit is governed by the Usadel equations~\cite{Buzdin_SFReview2005}. We find that the diffusive limit agrees best with our PdFe data in Fig.~\ref{fig:IcRn_Hshift_PdFe}, after fitting the points to the function,
\begin{equation}
\label{eq:IcRnZeroPiTransition}
I_c R_N = V_0*e^{-d_F/ \xi_{F1}}*\cos \left( \frac{d_F}{\xi_{F2}} - \phi \right).
\end{equation}
In Eqn.~\ref{eq:IcRnZeroPiTransition} $\xi_{F1}$ and $\xi_{F2}$ are length scales that control the decay and oscillation period of $I_c$ with $d_F$, and $\phi$ is an offset phase shift. $\xi_{F1}$, $\xi_{F2}$, and $\phi$ are used as fitting parameters. In diffusive systems, the simplest model of S/F/S Josephson junctions \cite{Buzdin_SFReview2005} predicts that $\xi_{F1} = \xi_{F2} = \sqrt{\hbar D_F/E_{ex}}$ and $\phi = \pi/4$, with $D_F$ and $E_{ex}$ being the diffusion constant and exchange energy of F, respectively. However, in cases with large spin-orbit or spin-flip scattering, one expects to find $\xi_{F1} < \xi_{F2}$ \cite{Faure2006}. Heim \textit{et al.}~\cite{Heim2015} have shown that the phase offset, $\phi$, varies sensitively with the thickness and type of normal-metal spacer layers or insulating barriers within the junction. The best-fit parameters are: $V_0= 85 \pm 13$ $\mu$V, $\xi_{F1}=13.6 \pm$ 1.3 nm, $\xi_{F2}= 6.71 \pm 0.37$ nm, and $\phi= 0.88 \pm 0.14$. The fits show that the junctions have $\xi_{F1} > \xi_{F2}$, which was also the case for a PdNi alloy studied previously~\cite{Khaire2009}. Bergeret \textit{et al.} have shown that $\xi_{F1} > \xi_{F2}$ is a persistent feature in the semi-clean limit where $\xi_{F1}=l_e$, the mean free path~\cite{Bergeret2001}. Our data suggest that Pd$_{97}$Fe$_{3}$ is also in the semi-clean limit.

In Fig.~\ref{fig:IcRn_Hshift_PdFe}(b), we plot the average of $H_{\mathrm{shift}}$ from the Fraunhofer pattern fits for each sweep direction versus $d_F$. We find that $H_{\mathrm{shift}}$ vs. $d_F$ increases due to the magnetic flux in the junction contributed by the uniform magnetization of the ferromagnet. Due to the fact that our $\lambda_L \gg d_F $ the trend is approximately linear. Despite the small magnetization of PdFe, large field shifts ($>15$ mT for the thickest samples measured), commensurate with the width of the central peak of the Fraunhofer pattern are observed due to the large PdFe thickness. We fit these data to:
\begin{equation}
\label{eqn:Hshift}
H_{\mathrm{shift}}=M (d_F-d_\mathrm{dead})/(2 \lambda_L+ 2 d_{\mathrm{Cu}}+ d_F),
\end{equation}
with $M$ and $d_\mathrm{dead}$ used as fitting parameters. The fit yields $M$ = 72 $\pm$ 16 kA/m and $d_\mathrm{dead}$ = -4 $\pm$ 5 nm. While the magnetization value obtained from Fig.~\ref{fig:IcRn_Hshift_PdFe}(b) lies within the uncertainty of that from Fig.~\ref{fig:PdFeSQUIDMagnetomemer}, the value of $d_\mathrm{dead}$ does not, even with it's large uncertainty. If we instead fix $d_\mathrm{dead}$=0 and re-fit the data in Fig.~\ref{fig:IcRn_Hshift_PdFe}(b) we find $M$ = 88 $\pm$ 4 kA/m, which is closer to the result from Fig.~\ref{fig:PdFeSQUIDMagnetomemer}.

\section{Conclusion}
In conclusion we have studied the magnetic and transport behavior of micron-scale SFS Josephson junctions containing Pd$_{97}$Fe$_3$. If used as a ``free'' magnetic layer in cryogenic memory, Pd$_{97}$Fe$_3$ is advantageous in that its 0-$\pi$ transition occurs at a thickness of $\approx$16.5 nm, much greater than for NiFe, making Pd$_{97}$Fe$_3$ much less sensitive to small thickness variations. Meanwhile, junctions with Pd$_{97}$Fe$_3$ maintain a relatively low switching field $|H_{\mathrm{switch,2}}|$ =  2.4 mT (with standard deviation 0.9 mT). As a ``free'' layer Pd$_{97}$Fe$_3$  has some disadvantages-- the magnetic switching can occur over a range of fields, possibly due to the existence of weakly coupled ferromagnetic clusters.  For reproducible magnetic switching, the junctions had to be magnetized at an initialization field of 5 mT or greater. In the future we plan to further increase the Fe concentration, in the range of 5-7 $\%$, to see if it is possible to improve the magnetic properties of this F-layer.

\section*{Acknowledgment}
 We thank B. Niedzielski, E. Gingrich, A. Herr, D. Miller, N. Newman, and N. Rizzo for helpful discussions, and B. Bi for help with fabrication using the Keck Microfabrication Facility. This research is supported by the Office of the Director of National Intelligence (ODNI), Intelligence Advanced Research Projects Activity (IARPA), via U.S. Army Research Office contract W911NF-14-C-0115. The views and conclusions contained herein are those of the authors and should not be interpreted as necessarily representing the official policies or endorsements, either expressed or implied, of the ODNI, IARPA, or the U.S. Government.

\ifCLASSOPTIONcaptionsoff
  \newpage
\fi



\bibliographystyle{IEEEtran}
\bibliography{IEEEabrv,Glick_SFS_PdFe}

\providecommand{\noopsort}[1]{} \providecommand{\singleletter}[1]{\#1}\%
\begin{thebibliography}{10}
\providecommand{\url}[1]{#1}
\csname url@samestyle\endcsname
\providecommand{\newblock}{\relax}
\providecommand{\bibinfo}[2]{#2}
\providecommand{\BIBentrySTDinterwordspacing}{\spaceskip=0pt\relax}
\providecommand{\BIBentryALTinterwordstretchfactor}{4}
\providecommand{\BIBentryALTinterwordspacing}{\spaceskip=\fontdimen2\font plus
\BIBentryALTinterwordstretchfactor\fontdimen3\font minus
  \fontdimen4\font\relax}
\providecommand{\BIBforeignlanguage}[2]{{%
\expandafter\ifx\csname l@#1\endcsname\relax
\typeout{** WARNING: IEEEtran.bst: No hyphenation pattern has been}%
\typeout{** loaded for the language `#1'. Using the pattern for}%
\typeout{** the default language instead.}%
\else
\language=\csname l@#1\endcsname
\fi
#2}}
\providecommand{\BIBdecl}{\relax}
\BIBdecl

\bibitem{Bell2004}
C.~Bell, G.~Burnell, C.~W. Leung, E.~J. Tarte, D.-J. Kang, and M.~G. Blamire,
  ``{Controllable Josephson current through a pseudospin-valve structure},''
  \emph{Appl. Phys. Lett.}, vol.~84, pp. 1153--1155, 2004.

\bibitem{Ryazanov2012}
V.~V. Ryazanov, V.~V. Bol’ginov, D.~S. Sobanin, I.~V. Vernik, S.~K. Tolpygo,
  A.~M. Kadin, and O.~A. Mukhanov, ``Magnetic josephson junction technology for
  digital and memory applications,'' \emph{Physics Procedia}, vol.~36, pp. 35
  -- 41, 2012.

\bibitem{Larkin2012}
T.~I. Larkin, V.~V. Bol’ginov, V.~S. Stolyarov, V.~V. Ryazanov, I.~V. Vernik,
  S.~K. Tolpygo, and O.~A. Mukhanov, ``Ferromagnetic josephson switching device
  with high characteristic voltage,'' \emph{Appl. Phys. Lett.}, vol. 100,
  no.~22, 2012.

\bibitem{Vernik2013}
I.~V. Vernik, V.~V. Bol'ginov, S.~V. Bakurskiy, A.~A. Golubov, M.~Y.
  Kupriyanov, V.~V. Ryazanov, and O.~A. Mukhanov, ``Magnetic josephson
  junctions with superconducting interlayer for cryogenic memory,'' \emph{IEEE
  Trans. App. Supercond.}, vol.~23, no.~3, p. 1701208, June 2013.

\bibitem{Qader2014}
M.~Abd El~Qader, R.~K. Singh, S.~N. Galvin, L.~Yu, J.~M. Rowell, and N.~Newman,
  ``Switching at small magnetic fields in josephson junctions fabricated with
  ferromagnetic barrier layers,'' \emph{Appl. Phys. Lett.}, vol. 104, no.~2,
  2014.

\bibitem{Baek2014}
B.~Baek, W.~H. Rippard, S.~P. Benz, S.~E. Russek, and P.~D. Dresselhaus,
  ``{Hybrid superconducting-magnetic memory device using competing order
  parameters},'' \emph{Nature Commun.}, vol.~5, p. 3888, 2014.

\bibitem{Gingrich2016}
E.~C. Gingrich, B.~M. Niedzielski, J.~A. Glick, Y.~Wang, D.~L. Miller,
  R.~Loloee, W.~P. {Pratt Jr}, and N.~O. Birge, ``{Controllable 0-$\pi$
  Josephson junctions containing a ferromagnetic spin valve},'' \emph{Nat.
  Phys.}, vol.~12, no.~6, pp. 564--567, Jun 2016.

\bibitem{Senoussi1977}
S.~Senoussi, I.~Campbell, and A.~Fert, ``Evidence for local orbital moments on
  $\mathrm{Ni}$ and $\mathrm{Co}$ impurities in $\mathrm{Pd}$,'' \emph{Solid
  State Communications}, vol.~21, no.~3, pp. 269 -- 271, 1977.

\bibitem{Arham2009}
H.~Z. Arham, T.~S. Khaire, R.~Loloee, W.~P. Pratt, and N.~O. Birge,
  ``Measurement of spin memory lengths in $\mathrm{PdNi}$ and $\mathrm{PdFe}$
  ferromagnetic alloys,'' \emph{Phys. Rev. B}, vol.~80, p. 174515, Nov 2009.

\bibitem{GingrichThesis2014}
E.~C. Gingrich, ``Phase control of the spin-triplet state in $\mathrm{S/F/S}$
  josephson junctions,'' Ph.D. dissertation, Michigan State University, 2014.

\bibitem{Wang2012}
Y.~Wang, W.~P. {Pratt Jr}, and N.~O. Birge, ``{Area-dependence of spin-triplet
  supercurrent in ferromagnetic Josephson junctions},'' \emph{Phys. Rev. B},
  vol.~85, p. 214522, 2012.

\bibitem{Thomas1998}
C.~D. Thomas, M.~P. Ulmer, and J.~B. Ketterson, ``Superconducting tunnel
  junction base electrode planarization,'' \emph{J. App. Phys.}, vol.~84,
  no.~1, pp. 364--367, 1998.

\bibitem{Kohlstedt1996}
H.~Kohlstedt, F.~König, P.~Henne, N.~Thyssen, and P.~Caputo, ``The role of
  surface roughness in the fabrication of stacked $\mathrm{Nb/Al–AlOx/Nb}$
  tunnel junctions,'' \emph{J. App. Phys.}, vol.~80, no.~9, pp. 5512--5514,
  1996.

\bibitem{Niedzielski2015}
B.~M. Niedzielski, E.~C. Gingrich, R.~Loloee, W.~P. Pratt, and N.~O. Birge,
  ``{S/F/S Josephson junctions with single-domain ferromagnets for memory
  applications},'' \emph{Supercond. Sci Technol.}, vol.~28, no.~8, p. 085012,
  2015.

\bibitem{Glick2016}
J.~A. Glick, M.~A. Khasawneh, B.~M. Niedzielski, E.~C. Gingrich, P.~G. Kotula,
  N.~Missert, R.~Loloee, W.~P.~J. Pratt, and N.~O. Birge, ``{Critical Current
  Oscillations of Elliptical Josephson Junctions with Single-Domain
  Ferromagnetic Layers},'' \emph{arXiv:1608.08998}, 2016.

\bibitem{Barone1982}
A.~Barone and G.~Patern{\`o}, \emph{{Physics and applications of the Josephson
  effect}}.\hskip 1em plus 0.5em minus 0.4em\relax Wiley, 1982.

\bibitem{Khaire2009}
T.~S. Khaire, W.~P. Pratt, and N.~O. Birge, ``Critical current behavior in
  $\mathrm{J}$osephson junctions with the weak ferromagnet $\mathrm{PdNi}$,''
  \emph{Phys. Rev. B}, vol.~79, p. 094523, Mar 2009.

\bibitem{Uspenskaya2013}
L.~S. Uspenskaya, A.~L. Rakhmanov, L.~A. Dorosinskii, A.~A. Chugunov, V.~S.
  Stolyarov, O.~V. Skryabina, and S.~V. Egorov, ``Magnetic patterns and flux
  pinning in $\mathrm{Pd_{0.99}Fe_{0.01}}$-$\mathrm{Nb}$ hybrid structures,''
  \emph{Pis'ma v Zhurnal Eksperimental'noi i Teoreticheskoi Fiziki}, vol.~97,
  no.~3, pp. 176--179, 2013, [\textit{J. Exp. Theor. Phys. Lett.}, \textbf{97},
  3, 155--158 (2013)].

\bibitem{Buzdin_SFReview2005}
A.~I. Buzdin, ``{Proximity effects in superconductor-ferromagnet
  heterostructures},'' \emph{Rev. Mod. Phys.}, vol.~77, pp. 935--976, Sep 2005.

\bibitem{Robinson2006}
J.~W.~A. Robinson, S.~Piano, G.~Burnell, C.~Bell, and M.~G. Blamire,
  ``{Critical current oscillations in strong ferromagnetic $\pi$ junctions},''
  \emph{Phys. Rev. Lett.}, vol.~97, p. 177003, 2006.

\bibitem{Robinson2007}
------, ``{Zero to $\pi$ transition in
  superconductor-ferromagnet-superconductor junctions},'' \emph{Phys. Rev. B},
  vol.~76, p. 094522, Sep 2007.

\bibitem{Faure2006}
M.~Faur{\'e}, A.~I. Buzdin, A.~A. Golubov, and M.~Y. Kupriyanov, ``{Properties
  of superconductor/ferromagnet structures with spin-dependent scattering},''
  \emph{Phys. Rev. B}, vol.~73, p. 064505, Feb 2006.

\bibitem{Heim2015}
D.~M. Heim, N.~G. Pugach, M.~Y. Kupriyanov, E.~Goldobin, D.~Koelle, R.~Kleiner,
  N.~Ruppelt, M.~Weides, and H.~Kohlstedt, ``{The effect of normal and
  insulating layers on 0-$\pi$ transitions in Josephson junctions with a
  ferromagnetic barrier},'' \emph{New J. Phys.}, vol.~17, no.~11, p. 113022,
  2015.

\bibitem{Bergeret2001}
F.~S. Bergeret, A.~F. Volkov, and K.~B. Efetov, ``Josephson current in
  superconductor-ferromagnet structures with a nonhomogeneous magnetization,''
  \emph{Phys. Rev. B}, vol.~64, p. 134506, Sep 2001.

\end{thebibliography}
\end{document}